# PUFFIN: A Path-Unifying Feed-Forward Interfaced Network for Vapor Pressure Prediction

Vinicius Viena Santana[a,b,c,*], Carine Menezes Rebello[a], Luana P. Queiroz[b,c], Ana Mafalda Ribeiro[b,c], Nadia Shardt[a,*] and Idelfonso B. R. Nogueira[a,*]

[a]*Department of Chemical Engineering; Norwegian University of Science and Technology, Gløshaugen, Trondheim, 7034, Norway*

[b]*LSRE-LCM - Laboratory of Separation and Reaction Engineering – Laboratory of Catalysis and Materials, Faculty of Engineering, University of Porto, Rua Dr. Roberto Frias, Porto, 4200-465, Portugal*

[c]*ALiCE - Associate Laboratory in Chemical Engineering, Faculty of Engineering, University of Porto, Rua Dr. Roberto Frias, Porto, 4200-465, Portugal*

ABSTRACT

Accurately predicting vapor pressure is vital for various industrial and environmental applications. However, obtaining accurate measurements for all compounds of interest is not possible due to the resource- and labor intensity of experiments. The demand for resources and labor further multiplies when a temperature-dependent relationship for predicting vapor pressure is desired. In this paper, we propose PUFFIN (Path-Unifying Feed-Forward Interfaced Network), a machine learning framework that combines transfer learning with a new inductive bias node inspired by domain knowledge (the Antoine equation) to improve vapor pressure prediction. By leveraging inductive bias and transfer learning using graph embeddings, PUFFIN outperforms alternative strategies that do not use inductive bias or that use generic descriptors of compounds. The framework's incorporation of domain-specific knowledge to overcome the limitation of poor data availability shows its potential for broader applications in chemical compound analysis, including the prediction of other physicochemical properties. Importantly, our proposed machine learning framework is partially interpretable, because the inductive Antoine node yields network-derived Antoine equation coefficients. It would then be possible to directly incorporate the obtained analytical expression in process design software for better prediction and control of processes occurring in industry and the environment.

## 1. Introduction

The prediction of vapor pressure, and particularly its temperature-dependent behavior, plays a central role in a range of industrial and environmental contexts. From refining operations in the petrochemical industry to projecting pollutant spread in environmental sciences, vapor pressure quantification serves as a foundational underpinning. However, due to the vast diversity and unique properties of chemical compounds, its accurate measurement remains an arduous and resource-intensive task, one that becomes more intractable when temperature-dependence is considered.

Traditional techniques necessitate not only specialized and often costly equipment but also time-consuming procedures to ensure the accuracy of vapor pressure measurements (Goll and Jurs, 1999). Furthermore, obtaining reliable experimental vapor pressure data for many compounds can be challenging, particularly for compounds with normal boiling points above 200 °C, due to a scarcity of data (Yaffe and Cohen, 2001). This reality presents a barrier to systematically exploring new applications for chemical molecules, particularly in high-throughput screening, a technique widely used in drug discovery and materials science. Added intricacies stem from compound availability, stability, and potential safety hazards. Therefore, there is a burgeoning interest in the development of alternative methodologies, specifically computational methods based on machine learning, to predict vapor pressure with a high degree of accuracy (Khosravi et al., 2018; Lansford et al., 2023).

Historically, a few methods have attempted to address these challenges. For instance, Tu (1994) leveraged a group contribution method for predicting vapor pressure at a fixed temperature. Later, Zang et al. (2017) introduced an open-source quantitative structure-property relationship (QSPR) workflow to predict various physicochemical properties, including vapor pressure. Their approach aimed to facilitate high-throughput screening by proposing a predictive model for thermodynamic properties that surpassed the accuracy of the original EPI Suite models for five out of the six predicted properties. This marked a significant step forward in demonstrating the potential of artificial intelligence (AI) tools in vapor pressure prediction. Furthermore, Zang et al. (2017) performed a statistical analysis, suggesting a strong correlation between the vapor pressure and the boiling point, evidence that will be further explored in this work in a transfer learning context. In Ohe (2019), a method was proposed to predict vapor pressure by directly determining the constants of the vapor pressure equation from boiling point data. The approach is rooted in the observed linear relationship between the logarithm of measured vapor pressure and the reciprocal value of absolute temperature across homologous series. Thus, by utilizing the slope of this relationship for a homologous substance, the intercept of the vapor pressure equation for a substance whose vapor pressure has not been measured can be estimated from its boiling point data.

*Corresponding authors
✉ up201700649@edu.fe.up.pt (V.V. Santana); nadia.shardt@ntnu.no (N. Shardt); idelfonso.b.d.r.nogueira@ntnu.no (I.B.R. Nogueira)
ORCID(s):

Accurately predicting sought-after properties, particularly in situations where data is limited, remains a significant hurdle across diverse scientific and engineering fields. In these contexts, transfer learning has emerged as an invaluable technique, given its capacity to repurpose a model engineered for one task as the foundation for another task (Sanchez-Lengeling et al., 2019; Queiroz et al., 2022). This approach has gained increasing interest where pre-trained models are usually used as the starting point for computer vision and natural language processing tasks (Zhuang et al., 2021; Vermeire and Green, 2021). In our specific context, applying transfer learning can be particularly beneficial when we aim to predict a property with limited data availability. Thus, the premise of transfer learning is grounded in the assumption that the features learned from the task with abundant data are relevant and useful for the task with limited data. This assumption holds especially well when the properties are physically related. For instance, the boiling point and vapor pressure of compounds are both linked to the strength of the intermolecular forces. Consequently, they could be expected to have related feature representations. We can leverage the abundant data for one related property to train an initial model and transfer the learned features and patterns to predict the property of interest.

Recent advancements in artificial intelligence and machine learning have further corroborated the capabilities of machine learning models for learning the relationship between molecular structures and properties at high dimensional data (Lansford et al., 2023; Fang et al., 2022), with graph neural networks (GNNs) emerging as a potent approach to tackle these challenges in QSPR modeling (Aouichaoui et al., 2023; Schweidtmann et al., 2020). Using graph theory to represent a compound as a graph, where each node is an atom, and each edge is a bond between atoms, GNNs capture the complex 3D structure of molecules, which is crucial for predicting their properties (Zhang et al., 2020; Ohe, 2019; Zhang et al., 2022; Queiroz et al., 2023a,b).

To further exploit these advancements, this paper introduces PUFFIN (Path-Unifying Feed-Forward Interfaced Network), a novel machine learning framework. PUFFIN integrates transfer learning and inductive bias to improve prediction accuracy and reliability. Specifically, transfer learning leverages the graph embeddings from a pre-trained GNN model. These graph embeddings, which encapsulate meaningful representations of molecules and their structural relationships (Sanchez-Lengeling et al., 2019), serve as inputs to a modified feed-forward neural network (FFNN) within the PUFFIN framework. This approach allows the FFNN to benefit from the GNN's understanding of complex relationships and patterns within the molecular graph data, thereby enhancing the accuracy of vapor pressure predictions. Additionally, PUFFIN incorporates inductive bias, ensuring the network output conforms to the functional form of known physical laws relevant to vapor pressure. This integration reinforces the model's capability to capture the underlying physical phenomena governing vapor pressure.

The introduction of PUFFIN marks a significant stride towards an effective methodology for predicting temperature-dependent vapor pressure rooted in a strong theoretical foundation. Combining the strengths of GNNs, transfer learning, domain knowledge, and inductive bias, PUFFIN presents a comprehensive framework for accurate and interpretable predictions. By ensuring alignment with known physical laws through an internal hybridization of the neural network structure, PUFFIN not only enriches the existing array of computational models but also fortifies the bridge between machine learning and domain-specific knowledge of physical phenomena. This holds promising implications for the future of scientific research and practical applications in industries reliant on accurate vapor pressure prediction.

## 2. Methodology

The methodology section in this paper focuses on implementing three distinct methodological paths, each of which plays a crucial role in the overall analysis. Figure 1 serves as a visual representation of these paths, with each path being highlighted in the figure using its respective color. The first two paths are introduced as benchmark tests, serving as reference points for comparison with the novel approach proposed in the paper. The subsequent subsections of the methodology will delve into detailed descriptions of each path, providing a comprehensive understanding of their individual contributions and how they relate to the central focus of this research.

In Section 2.1, we present a methodology that focuses on creating molecular descriptors based on a specific target property. To achieve this, we leverage the power of graph neural networks, which allow us to find a mathematical representation of the molecules guided by the desired property. This approach enables us to capture the intricate relationships and structural patterns within the molecules. The hypothesis underlying this approach is that it can create more accurate and informative descriptors. This hypothesis will be tested against a generic descriptor.

Moving on to Section 2.2, we introduce an alternative approach that uses a generic descriptor of the molecules. This descriptor utilizes logical rules to obtain a numerical molecular representation in the form of a binary vector. This is in contrast to the embedding, which is a matrix representation between two neural networks. By comparing these two descriptor types, we can assess the potential benefits of utilizing graph descriptors that are obtained from a target property (such as boiling point).

Lastly, in Section 2.3, we present our main proposal, which centers around integrating graph descriptors and knowledge reinforcement in a single framework. This novel methodology aims to enhance the predictive capabilities of neural networks in relation to a specific property. By leveraging the unique strengths of graph descriptors and incorporating knowledge reinforcement techniques, we strive to develop a robust and accurate neural network model for property prediction.

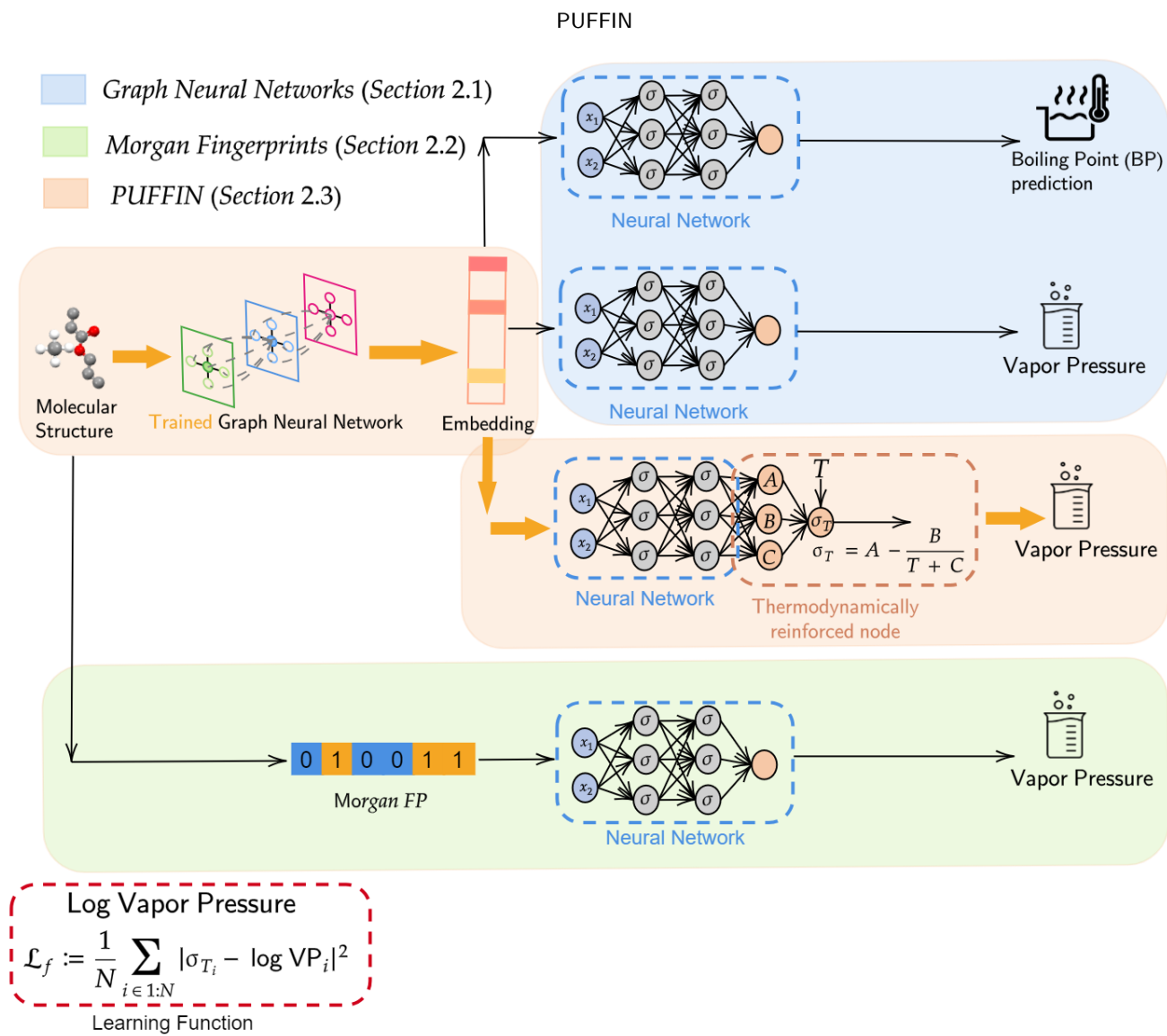

**Figure 1**: Schematic diagram of the methodology presented in this work

## 2.1. Graph Neural Networks and Transfer Learning

In this study, we have employed a two-tier approach to model and predict the vapor pressure of chemical compounds. The first stage utilizes a graph neural network to predict the boiling point of the compounds, a property typically more accessible in data terms than vapor pressure. This GNN then serves as a benchmark model whose generated embeddings are transferred to a feed-forward neural network (FFNN) to predict the vapor pressure. The principal idea is to evaluate the accuracy of our proposed machine learning framework—the Path-Unifying Feed-Forward Interfaced Network—compared to the two-stage GNN and FFNN model. The transfer learning principle presented here will also be used within the PUFFIN framework. Figure 1 provides a concise visual depiction of the comprehensive framework underlying this approach, highlighted in blue.

A GNN is well-suited for the task of representing molecules as it excels at handling graph-structured data. This hypothesis will be verified in this paper by comparing the GNN results with the outcome obtained by employing a generic descriptor (the Morgan fingerprint). In this context, the molecules of the compounds are modeled as graphs where atoms are nodes and chemical bonds are edges. These GNNs are capable of capturing the local and global relationships between atoms in a molecule, which is essential for the accurate prediction of physicochemical properties. Mathematically, a GNN works by iteratively updating the representations (or embeddings) of each node based on the representations of its neighbors, following a procedure often referred to as message passing or neighborhood aggregation:

$$h_i^{(l+1)} = f_h^{(l)}\left(h_i^{(l)}, \text{Aggregate}^{(l)}\left(\{h_j^{(l)} : j \in N(i)\}\right)\right) \quad (1)$$

Here, $h_i^{(l)}$ denotes the feature vector of node $i$ at layer $l$, $N(i)$ is the set of its neighboring nodes, and $f_h^{(l)}$ and $Aggregate^{(l)}$ are parameterized functions implemented as neural networks. This process is repeated for a fixed number of iterations (or layers), after which the final node representations are used to make the prediction.

Within the transfer learning framework, the generated embeddings from the GNN model serve as inputs to the FFNN model, allowing it to leverage the GNN's understanding of complex relationships and patterns in the molecular graph data (Zang et al., 2017). FFNNs are simple, yet powerful, neural networks that map inputs to outputs through a series of transformations across multiple layers, each followed by a non-linear activation function. The mathematical formulation of a layer in an FFNN is as follows:

$$h^{(l+1)} = f_h^{(l)}(W^{(l)}h^{(l)} + b^{(l)}) \quad (2)$$

where $h^{(l)}$ denotes the activations of layer $l$, $W^{(l)}$ and $b^{(l)}$ are the weight matrix and bias vector of layer $l$, respectively, and $f_h^{(l)}$ is the activation function of layer $l$. As indicated by the preceding equation, the inputs of the FFNN undergo internal processing to orient toward a new target. In the current context, the GNN-derived embeddings, originally intended for boiling point prediction, are processed by the FFNN to predict vapor pressure. This methodology is backed by evidence in scientific literature pointing to the boiling point as a critical and insightful variable in transfer learning applications aiming at vapor pressure prediction Zang et al. (2017); Ohe (2019). This established correlation between boiling point and vapor pressure reinforces the potential of transfer learning.

## 2.2. Generic Descriptor: Morgan Fingerprints

In our methodology, we also consider another benchmark strategy for evaluating the performance of the PUFFIN framework. This alternative approach bypasses the use of inductive bias and transfer learning, instead directly predicting the vapor pressure of compounds. At the heart of this strategy is using Morgan's fingerprints as generic molecular descriptors (Morgan, 1965). These fingerprints provide a numerical representation of molecules and offer an alternative to the GNN used in our benchmark model and within the PUFFIN framework. The Morgan algorithm starts by assigning each atom in the molecule an initial identifier (usually its atomic number). Then, for each iteration up to a specified radius, it updates the identifier for each atom to a hash of its own identifier and the identifiers of its immediate neighbors. Finally, the identifiers are hashed into the fingerprint bit vector. Figure 1 provides a concise visual depiction of the comprehensive framework underlying this approach, highlighted in green.

Mathematically, a molecule's Morgan fingerprint, **F**, is a binary vector where each element, $f_i$, represents the presence (1) or absence (0) of a specific substructure in the molecule. For a molecule with $n$ atoms, the fingerprint is computed as follows:

$$\mathbf{F} = [f_1, f_2, ..., f_n] \quad (3)$$

where $f_i$ is computed based on the connectivity and types of atoms in a circular neighborhood with a given radius around the $i^{th}$ atom.

In more detail, the $f_i$ is determined as:

$$f_i = \begin{cases} 1 & \text{if specific substructure exists around the } i^{th} \text{ atom,} \\ 0 & \text{otherwise.} \end{cases} \quad (4)$$

In this strategy, each molecule is represented by its Morgan fingerprint, which serves as input to a feed-forward neural network (FFNN). The FFNN, as a universal function approximator, aims to learn a mapping from these fingerprints to the vapor pressure, similar to our two other approaches.

This benchmark strategy not only serves as a comparative measure of performance but also highlights the efficacy of the PUFFIN framework and illuminates the enhancements gained through the transfer learning over using a direct numerical molecular representation like Morgan fingerprints.

## 2.3. PUFFIN: Path-Unifying Feed-Forward Interfaced Network

Our proposed PUFFIN framework enhances this two-stage model by integrating domain-knowledge reinforcement via inductive bias, ensuring alignment with known physical laws. The ultimate aim of this methodology is to develop a robust, efficient, and accurate tool for predicting vapor pressure that extends beyond the capabilities of conventional GNNs and FFNNs. Figure 1 provides a concise visual depiction of the comprehensive framework underlying this approach, highlighted in salmon.

The Path-Unifying Feed-Forward Interfaced Network incorporates an inductive bias node—an Antoine equation layer—into the traditional FFNN model, thereby reinforcing the phenomena within the model and offering a more accurate prediction of the vapor pressure. The Antoine equation is a well-established empirical model for describing the relationship between vapor pressure and temperature for a given substance, often expressed as $\log P = A - \frac{B}{(C+T)}$ (Thomson, 1946), where $P$ represents vapor pressure, $T$ is the temperature, and $A$, $B$, and $C$ are substance-specific parameters. Similar to the GNN-based approach, PUFFIN begins by learning embeddings of molecules using a graph neural network. These embeddings encapsulate structural and chemical information about the molecules and serve as inputs to the FFNN. The modified FFNN operates as described in Figure 1, where $\sigma_T$ is the output of the Antoine neuron, $A$, $B$ and $C$ are the outputs of the intermediary layer, and $T$ is the temperature.

In this hybridized FFNN, the Antoine equation serves as the activation function of the output layer. This is an innovative approach that exploits both the neural networks and the domain-specific knowledge encapsulated in the Antoine equation. The bias of this output neuron is considered an exogenous variable, the temperature ($T$), while its inputs are the outputs of the previous layer. These inputs, therefore, are effectively equivalent to the Antoine coefficients. This design essentially bridges the gap between the high-dimensional latent space represented by the previous layer's outputs and the real-world physical property to be predicted, the vapor pressure. The output of the Antoine neuron is the predicted vapor pressure, directly incorporating the temperature dependency encapsulated in the Antoine equation. This predicted vapor pressure is then compared with the true value to compute the loss function, which guides the training of the neural network.

By assessing PUFFIN's performance against the two benchmark models described in the subsections above, we seek to substantiate the advantages of incorporating transfer learning and inductive bias into machine learning models for predicting physical properties. This comparison demonstrates the effectiveness of PUFFIN's unique design, thereby showcasing its potential in advancing the field of computational prediction of physicochemical properties.

### 2.4. Vapor Pressure Data Set

The database used in this study consists of 1851 molecules of organic chemicals, encompassing a wide range of classes such as industrial chemicals, antimicrobials, dyes, fertilizers, flame retardants, fragrances, pharmaceutical products, herbicides, pesticides, inert ingredients, petrochemicals, food additives. The Simplified Molecular Identification and Line Entry System (SMILES) system represents these molecules in the database (Weininger, 1988). SMILES employs an alphanumeric character sequence to describe the connectivity of atoms in a molecule. Atoms are represented by their chemical symbols, and their bonds are indicated by special characters such as hyphens and numbers. Additionally, SMILES can incorporate information about functional groups, isomerism, stereochemistry, and electric charge.

For each molecule, vapor pressures were extracted at five different temperatures (298 K, 273 K, 325 K, 350 K, 375 K) to populate the database. The vapor pressures at 298 K were obtained experimentally by Zang et al. (2017), and the vapor pressures at the remaining temperatures were obtained through the "Thermo: Thermodynamics and Phase Equilibrium component of Chemical Engineering Design Library (ChEDL)" API (Bell and Contributors, 2023).

The training, validation, and test sets were obtained by shuffling the vapor pressure (VP) data set and first splitting it into two data sets—training and validation (80%) and test (20%). The training and validation were then split into two again, resulting in the training set (64%) and validation set (16%).

### 2.5. Hyperparameter Identification

Hyperparameter optimization is an essential part of any machine learning project, and in the development of the PUFFIN model, as well as the benchmark models, we have employed a state-of-the-art hyperparameter optimization method called HYPERBAND (Li et al. (2018)) to identify the optimal hyperparameters. Ensuring that the same method of optimization is used across all models allows for a fair and unbiased comparison of their performance.

The HYPERBAND algorithm, introduced by Li et al. (2018), is an efficient random search algorithm designed for hyperparameter optimization. It tackles the fundamental challenge of allocating limited computational resources to the evaluation of numerous hyperparameter configurations. Unlike conventional hyperparameter tuning methods that follow a sequential process, HYPERBAND employs a more dynamic approach, enabling it to explore the hyperparameter space more efficiently.

HYPERBAND operates on the principle of adaptive resource allocation and early-stopping. It begins by sampling a large number of hyperparameter configurations and allocating a small amount of resources to each. After an initial evaluation, only a fraction of these configurations, specifically the ones showing the most promise, are retained and given additional resources for further evaluation. Mathematically, HYPERBAND employs a multi-armed bandit strategy known as "successive halving" to allocate resources. If $n$ is the number of configurations and $r$ is the initial amount of resources allocated to each configuration, the algorithm iteratively reduces $n$ by a factor of $\mu$ (usually set to 3) while increasing $r$ by the same factor until only the most promising configuration is left. This is done over a series of $s$ brackets, each representing a round of resource allocation. By doing this, HYPERBAND eliminates sub-optimal configurations early in the process, allowing it to dedicate more resources to evaluating promising configurations. As a result, it can effectively identify the optimal hyperparameters within a relatively short time and with limited computational resources.

## 3. Results

### 3.1. Morgan Fingerprints—Vapor Pressure Prediction

Our baseline model was constructed as a feed-forward neural network taking inputs from Morgan fingerprints combined with temperature data. The output from this model was the base-10 logarithm of vapor pressure. To determine the optimal neural network architecture, we used HYPERBAND for hyperparameter search. We allowed for a range in the number of layers (from 1 to 6) and varied the number of neurons per layer (from 20 to 440 with increments of 40 neurons). Additionally, the learning rate was set to vary between $1 \times 10^{-5}$ and $1 \times 10^{-2}$.

To measure the effectiveness of these variations, we used a validation set to compute the mean absolute error, which served as our optimization metric. The training process was conducted with the Adaptive Moment Estimation (ADAM), (Kingma and Ba, 2017) algorithm, batching data in fixed sizes of 64. HYPERBAND was configured to a maximum of 100 epochs and a factor of 3 for the training process. The best architecture, identified by achieving the smallest validation set error, is presented in Table 1. Upon establishing the optimal architecture, the model underwent further training for 600 epochs, with an early stopping protocol in place to mitigate the risk of overfitting. The resultant mean absolute errors for the training and validation sets were 0.1663 and 0.3529, respectively.

Table 1
Architecture summary of identified feed forward neural network for predicting vapor pressure from Morgan fingerprints.

| Layer Type | Units | Activation |
|---|---|---|
| Input | 1024 | - |
| Dense | 340 | reLU |
| Dense | 60 | reLU |
| Dense | 100 | reLU |
| Dense | 1 | linear |

Table 2
Architecture summary of identified graph neural network (GNN) for predicting boiling point.

| Layer Type | Units | Activation |
|---|---|---|
| GNN | 128 | reLU |
| GNN | 64 | reLU |
| GNN | 64 | reLU |
| Global Add | - | - |
| Dense | 128 | reLU |
| Dropout | 0.2 probability | - |
| Dense | 128 | reLU |
| Dropout | 0.2 probability | - |
| Dense | 1 | linear |

To evaluate the baseline model's efficacy, the error against a test set was computed. This test dataset had been excluded from both the initial training and the hyperparameter tuning stages. The mean absolute error determined for this test set was 0.3690. To demonstrate the model's predictive proficiency with the test set, Figure 3.a illustrates a parity plot and a kernel density estimation of the residuals' distribution, showing that the mean of the residuals is centered around zero with low variance.

### 3.2. Molecular Embedding—Estimation From Boiling Point

Morgan fingerprints are a form of generic molecular descriptors that serve to encode molecular structures into a compact vector form. While these descriptors can provide a useful overview of molecular properties, they may not be the most suitable for predicting specific properties like vapor pressure due to their generic nature. An alternative approach could involve designing a more specialized descriptor for the task at hand. For instance, following the methodology presented in the previous section, we trained a separate neural network to predict the boiling point of a compound. The intermediate representation or "latent space" (also known as the embedding) from this boiling point prediction model is then used as a more specialized descriptor for predicting vapor pressure. Unlike traditional neural networks, a GNN is specifically designed to work with graph-like data structures, such as molecular structures, by considering their topological information. The input to the GNN was the molecular graph. During the training process, the GNN learned to transform these molecular graphs into a continuous vector representation, known as an embedding, which effectively captured the essential structural features relevant to the boiling point.

The architecture of our model consisted of three GNN layers. These were followed by a global addition pooling layer, two dense layers with rectified linear unit (ReLU) activation functions, and dropout layers interspersed for regularization. A final readout layer completed the architecture. Before proceeding with model identification, data preprocessing steps were undertaken. DeepChem library (Ramsundar et al., 2019) was used to convert the molecular SMILES strings into graph data. Subsequently, the target variable (boiling point) was standardized; this was achieved by subtracting the mean and dividing by the standard deviation of the boiling points in the training set.

The boiling point (BP) dataset was then partitioned into training and validation subsets, which constituted 80% of the total dataset. These subsets were used for hyperparameter tuning and for scaling operations. The remaining 20% of the dataset was designated as the test set and was reserved exclusively for the final evaluation of model performance. Hyperparameters such as the number of neurons per layer, batch size, and dropout probability were tuned using the ASHAScheduler (AsyncHYPERBANDScheduler). Specifically, the search space for the number of neurons per layer was defined as $2^i$, where $i$ could be any value from the set 6, 7, 8, 9. Dropout probability was allowed to vary among 0.2, 0.3, 0.4, providing a balance between model complexity and risk of overfitting. The learning rate was searched in a loguniform space, ranging from $5\times10^{-4}$ to $1\times10^{-2}$. Finally, the batch size was varied among the set 32, 64, 128.

The optimal model architecture determined from the hyperparameter search is detailed in Table 2. Upon establishing the optimal architecture, the model underwent further training for 300 epochs, with an early stopping protocol to mitigate the risk of overfitting. The final mean squared error for the test set with respect to the rescaled boiling point was 0.0471.

Figure 2 displays a parity plot showing the performance of the identified model on BP test set data. It can be seen that the data points are randomly spread over the identity line, which indicates satisfactory predictive performance.

The learned embeddings serve as a more specialized form of the molecular descriptor and learn a representation of the molecular structures in relation to their boiling point, which is intended to be used as input descriptors instead of the more generic Morgan fingerprints for predicting vapor pressure.

### 3.3. Molecular Embedding Without Domain Knowledge Node—Vapor Pressure Prediction

As mentioned, the 128-dimensional learned embeddings are intended to serve as a more specialized form of the molecular descriptor and learn a representation of the molecular structures in relation to their boiling point. As in the baseline model, to determine the optimal neural network architecture, we used HYPERBAND for hyperparameter search. We allowed for a range in the number of layers (from 1 to 6) and varied the number of neurons per layer (from 20 to 440 with increments of 40 neurons). Additionally, the learning rate was set to vary between $1 \times 10^{-5}$ and $1 \times 10^{-2}$.

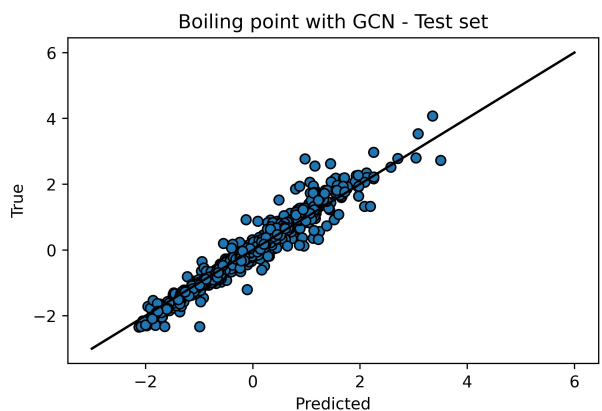

**Figure 2:** Parity plot of graph neural network (GNN) prediction of boiling point compared to the ground truth in the test set

**Table 3**
Architecture summary of the identified feed forward neural network for predicting vapor pressure from embeddings.

| Layer Type | Units | Activation |
|---|---|---|
| Dense | 260 | reLU |
| Dense | 220 | reLU |
| Dense | 180 | reLU |
| Dense | 1 | linear |

Table 3 presents the architecture of the identified feed-forward neural network used to predict vapor pressure at different temperatures from the learned embeddings.

Upon establishing the optimal architecture, the model underwent further training for 600 epochs, with an early stopping protocol to mitigate the risk of overfitting. The resultant mean absolute errors for the training and validation sets were measured at 0.2016 and 0.3646, respectively.

Figure 3.b shows a visual illustration of the model's predictive performance in the test set.

As can be seen from test set errors and plots, using embeddings from boiling point GNN predictive model did not produce a better predictive performance for the vapor pressure prediction task than the Morgan fingerprint generic descriptor.

### 3.4. Molecular Embedding With Domain Knowledge Node (PUFFIN)—Vapor Pressure Prediction

Finally, we evaluated the performance of the PUFFIN. To determine the optimal neural network architecture, we used HYPERBAND for hyperparameter search and allowed the same search space as for the other approaches. Table 4 presents the architecture of the identified feed-forward neural network used to predict vapor pressure at different temperatures from the learned embeddings.

**Table 4**
Architecture summary of the identified feed forward neural network for predicting vapor pressure from embeddings with domain knowledge reinforcement (the Antoine node).

| Layer Type | Units | Activation |
|---|---|---|
| Dense | 180 | reLU |
| Dense | 380 | reLU |
| Dense | 180 | reLU |
| Dense | 420 | reLU |
| Dense | 380 | reLU |
| Dense | 380 | reLU |
| Dense | 3 | linear |

Upon establishing the optimal architecture, the model underwent further training for 600 epochs, with an early stopping protocol in place to mitigate the risk of overfitting. The resultant mean absolute errors for the training and validation sets were measured at 0.1227 and 0.1609, respectively.

Figure 3.c illustrates the model's predictive performance in the test set. As can be seen both from the test set error and from the figures, the performance of PUFFIN reduced the error by about 50%. The residuals are considerably more concentrated around zero and with smaller variances than other approaches. This improvement can be attributed to the inductive bias introduced by using the Antoine equation in an intermediary node of the neural network. Within the proposed framework, we encode molecular information and boiling point data into the domain knowledge-reinforced PUFFIN, allowing a considerably better model for predicting vapor pressure.

Table 5 offers a summary of the test set performance for three different approaches: using Morgan fingerprints, using boiling point specialized embeddings, and using boiling point specialized embeddings with thermodynamic reinforcement. The performance is measured in terms of mean squared error (MSE) on the test set.

From the given results, it is evident that the approach using Morgan fingerprints has the worst performance, with an MSE of 0.3690. The performance improves slightly when the model uses boiling point specialized embeddings as the input, resulting in an MSE of 0.3646. This provides evidence for the hypothesis underlying this approach: guided graphs can create more accurate and informative descriptors. However, this hypothesis should be further explored in future work to understand its benefits and drawbacks.

The most significant improvement is observed when boiling point specialized embeddings are combined with domain knowledge reinforcement—in this case, the Antoine equation. This approach results in an MSE of 0.1609, more than halving the error obtained using the other two methods.

This result illustrates the utility of integrating domain-specific knowledge (in this case, boiling point) and physical laws (thermodynamic reinforcement) into the learning process, leading to enhanced prediction accuracy. These results underscore the potential for specialized embeddings and physics-informed machine learning to be used as predictive models for complex properties like vapor pressure.

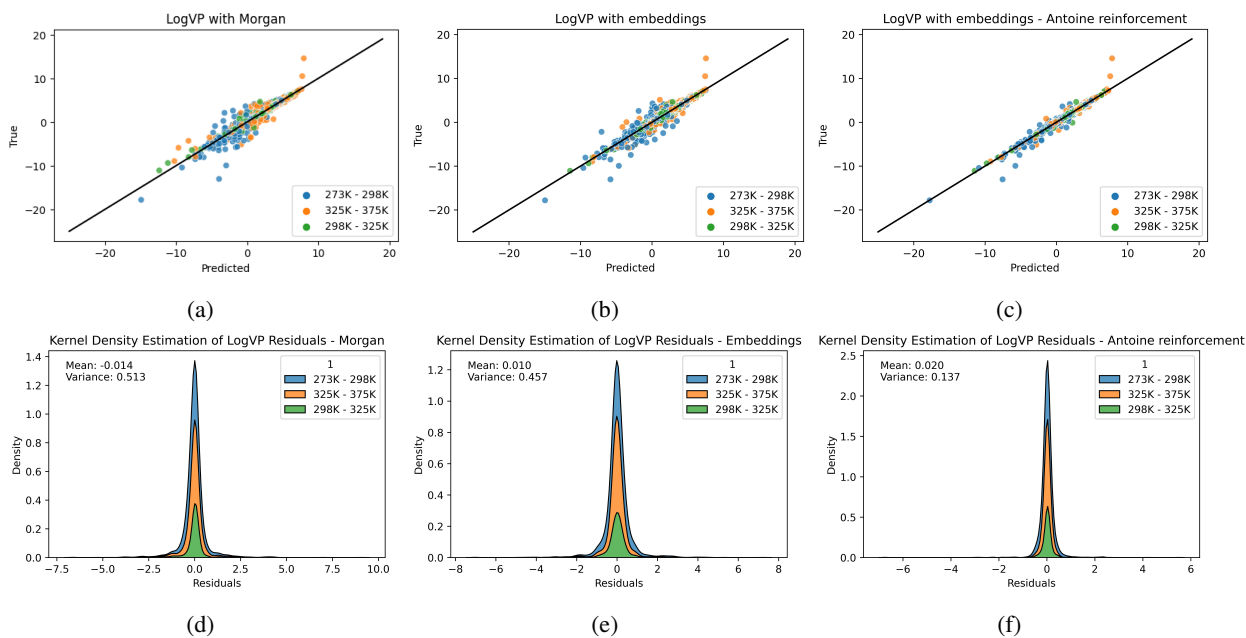

**Figure 3:** Comparative analysis: parity plots and kernel density estimation of residuals, including mean and variance for each model in the test dataset.

**Table 5**
Summary of test set performance for the three approaches evaluated based on Mean Squared Error (MSE) metric.

|  | Morgan fingerprint | Embeddings without Antoine | Embeddings with Antoine (PUFFIN) |
|---|---|---|---|
| Test set mse | 0.3690 | 0.3646 | 0.1609 |

**Table 6**
Comparison of coefficients predicted from PUFFIN and from literature.

| Substance | A_Puffin | A | B_Puffin | B | C_Puffin | C |
|---|---|---|---|---|---|---|
| Acetone | 9.019 | 12.312 | 1082.27 | 1315.67 | -46.31 | -32.671 |
| Butanol | 10.064 | 12.3013 | 1680.87 | 1285.0227 | -59.25 | -99.903 |
| Octyl-benzene | 9.402 | 12.46418 | 2173.842 | 2121.3576 | -63.591 | -71.162 |

### 3.5. Interpretability remarks

The intermediary layer of the PUFFIN model, which feeds into the Antoine neuron, produces output parameters $A$, $B$, and $C$. To test the partial interpretability of this structure, these parameters are evaluated against the known Antoine coefficients available at Dortmund Data Bank (www.ddbst.com) for three different components from the test set. Table 6 presents the values of the coefficients for three components and the embedding values provided by the PUFFIN.

The close agreement between these sets of coefficients underscores the model's ability to extract meaningful, domain-specific features from high-dimensional molecular graph data. For all three components, the model's $A$, $B$, and $C$ parameters closely align with the empirical Antoine coefficients. The accuracy of these outputs reinforces the model's capability to integrate the inductive bias of the Antoine equation into the predictive framework, ensuring not only high predictive accuracy but also thermodynamic consistency.

It is worth highlighting that the model was not fed any information about the Antoine coefficients during its training. Despite this, the PUFFIN model demonstrated a remarkable ability to generate embeddings consistent with the Antoine equation. The findings underscore the robustness of the PUFFIN model, given that it has been trained purely on high-dimensional molecular data and temperature–vapor pressure pairs, with no explicit information regarding the Antoine coefficients being provided.

These findings suggest that the hybridized structure of PUFFIN provides a methodology for incorporating domain-specific knowledge into the machine learning architecture, which can be extended to other fields. This mechanism promotes a richer interpretation of the model's outputs and contributes to its performance in accurately predicting vapor pressure. In particular, the successful retrieval of Antoine-like coefficients from the intermediary layer validates the design choice to use the Antoine equation as an activation function for the output layer. This lends credence to the concept that incorporating domain knowledge into machine learning models can improve both the interpretability and the predictive performance of these models.

A final assessment is provided in Figure 4. The figure presented in this analysis illustrates the extrapolation of three strategies in predicting vapor pressure, specifically focusing on the proposed strategy, PUFFIN. Figure 4 demonstrates the performance of each strategy relative to Antoine predictions, particularly in regions extending beyond the confines of the training data.

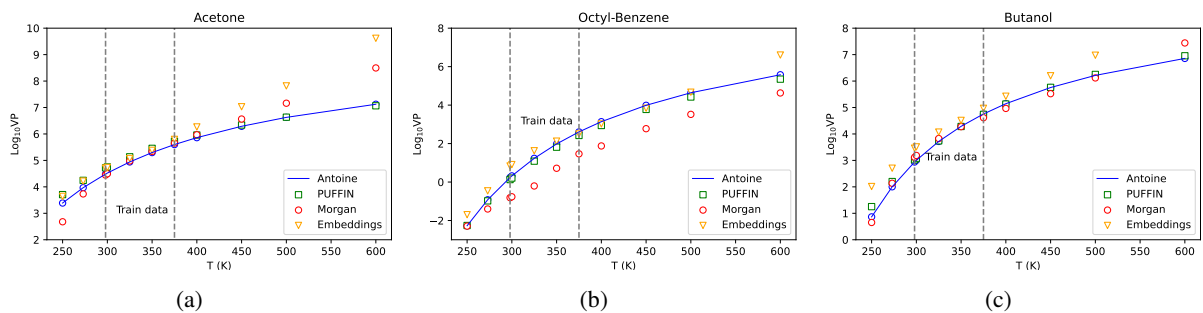

**Figure 4:** Extrapolation assessment of all models for three test set components. (a) Acetone, (b) Octyl-Benzene and (c) Butanol.

Notably, the PUFFIN strategy aligns accurately with Antoine predictions, even in regions significantly distant from the original training data where the other models drift significantly, a behavior clearly seen in Figure 4.a. This observation is important as it highlights the potential of our modified AI model to predict vapor pressure as a continuous function of temperature with remarkable accuracy.

In contrast, the other two strategies start to diverge noticeably from Antoine predictions when venturing into regions distant from the training area. This stark difference further emphasizes the exceptional precision and robustness of the PUFFIN strategy. Furthermore, it is essential to underline that Antoine coefficients are typically derived from experimental data, making it a significant challenge to obtain these values for all the molecules incorporated in our database. Despite this, our figure evaluates three molecules from the unseen test dataset.

The results from these evaluations show that PUFFIN can accurately predict vapor pressure for molecules unseen during the training phase and for temperatures well beyond the range of the training set. This capability highlights the potential of PUFFIN in predicting vapor pressure for new molecules and for temperatures outside the known training data, making it a compelling area for future research.

## 4. Conclusions

In conclusion, this study presents a novel approach for predicting vapor pressure using a hybrid neural network model, the PUFFIN framework, demonstrating good accuracy and partial interpretability. We adopted a transfer learning strategy, leveraging widely available boiling point data. Furthermore, we incorporated domain knowledge (thermodynamic) concepts into the neural network architecture with an inductive bias. Our results indicate that this approach significantly enhances the model's prediction capabilities, reducing the mean squared error by about 50% compared to the other two neural network approaches (the Morgan fingerprint network and the non-reinforced molecular embedding network).

Moreover, the PUFFIN framework's ability to extract Antoine-like coefficients directly from high-dimensional molecular graph data offers a compelling illustration of the benefits of incorporating domain knowledge into machine learning models. This promising outcome offers a degree of model interpretability typically lacking in standard neural network approaches, contributing significantly to the model's utility. Remarkably, these advances were achieved without providing the model with explicit information about the Antoine coefficients during training. This demonstrates the power of machine learning models to learn and utilize domain-specific relationships effectively, even without explicit domain knowledge during training.

Furthermore, the introduction of an Antoine node in the PUFFIN permitted a physically-accurate extrapolation of vapor pressure data to temperatures beyond those to which training was completed. This is in contrast to the two other methods (the Morgan fingerprint and the non-reinforced molecular embedding), which were both less accurate in predicting vapor pressure at temperatures outside of the range used for training.

The improvements realized by the PUFFIN framework affirm the considerable potential of incorporating domain-specific knowledge into machine learning architectures. The demonstrated success of this approach suggests potential applicability beyond vapor pressure prediction, extending to other domains that could benefit from a similar integration of domain knowledge into machine learning models. In the face of ever-increasing complexity and data availability in various scientific and engineering domains, our study underscores the importance and potential of physics-informed and interpretable machine learning models for accurately predicting complex properties. This further underscores the potential of the PUFFIN model as a reliable instrument for predicting complex thermophysical properties.

## 5. Acknowledgements

This research was supported by the doctoral Grant (reference PRT/BD/152850/2021) with funds from State Budgets under MIT Portugal Program. This work was also financially supported by LA/P/0045/2020 (ALiCE), UIDB/50020/2020 and UIDP/50020/2020 (LSRE-LCM), funded by national funds through FCT/MCTES (PIDDAC).

## 6. Declaration of generative AI and AI-assisted technologies in the writing process

During the preparation of this work the authors used ChatGPT 4.0 in order to improve language and readability. After using this tool, the authors reviewed and edited the content as needed and take full responsibility for the content of the publication.